\documentclass[fleqn,twoside]{article}
\usepackage{espcrc2}
\usepackage{epsfig}
\usepackage{amssymb}
\usepackage{color,epsfig,graphicx,longtable}

\usepackage{graphicx}
\usepackage[figuresright]{rotating}

\newcommand{\AmS}{{\protect\the\textfont2
  A\kern-.1667em\lower.5ex\hbox{M}\kern-.125emS}}

\newcommand{\beqn}{\begin{eqnarray}}
\newcommand{\eeqn}{\end{eqnarray}}

\newcommand{\Dlr}{\stackrel{\leftrightarrow}{D}}
\newcommand{\Dl}{\stackrel{\leftarrow}{D}}
\newcommand{\Dr}{\stackrel{\rightarrow}{D}}

\newcommand{\gRcf}{\frac{g_R^2 \, C_F}{16 \pi^2} }
\newcommand{\gTIcf}{\frac{g_{TI}^2 \, C_F}{16 \pi^2} }

\newcommand{\cO}{\mathcal {O}}

\newcommand{\MS}{\overline{MS}}
\newcommand{\csw}{c_{sw}}

\title{
\thispagestyle{empty}
\vspace{-17mm}
\rightline
{\small DESY 05-228, Edinburgh 2005/20, LU-ITP 2005/023, LTH 683}
\vspace{12mm}
Perturbative Renormalisation for Low
Moments of Generalised Parton Distributions
with Clover Fermions
}

\author{M.~G\"ockeler\address[MCSD1]{Institut f\"ur Theoretische Physik, Universit\"at
                          Regensburg, 93040 Regensburg, Germany},
        R.~Horsley\address{School of Physics, University of Edinburgh, Edinburgh EH9 3JZ, UK},
        H.~Perlt\address[ITPL]{Institut f\"ur Theoretische Physik, Universit\"at
			Leipzig, 04109 Leipzig, Germany},
        P.~E.~L.~Rakow\address{Theoretical Physics Division, Department of Mathematical Sciences,
          University of Liverpool,\\ Liverpool L69 3BX, UK},
        A.~Sch\"afer\addressmark[MCSD1],
        G.~Schierholz\address{John von Neumann-Institut f\"ur Computing NIC,
         Deutsches Elektronen-Synchrotron DESY, 15738 Zeuthen, Germany}%
		\address{Deutsches Elektronen-Synchrotron DESY, 22603 Hamburg, Germany},
        and
        A.~Schiller\addressmark[ITPL]\thanks{Talk given by A. Schiller at the Workshop on Computational 
	            Hadron Physics, Nicosia, September 2005.}}
       
\begin{document}

\begin{abstract}
We present the non-forward quark matrix elements of operators with one and two
  covariant derivatives needed for the renormalisation of the first and second moments
  of generalised parton distributions in one-loop lattice perturbation theory
  using clover fermions. For some representations of the hypercubic group
  commonly used in simulations we define the sets of possible mixing
  operators and compute the one-loop mixing matrices
  of renormalisation factors. Tadpole improvement is applied to the results 
  and some numerical examples are presented.
\vspace{1pc}
\end{abstract}

\maketitle

\section{INTRODUCTION}

Generalised parton distributions (GPDs)
have become a focus of both experimental and theoretical studies in hadron physics
(for an extensive up-to-date review  see \cite{Diehl}). They
allow a parametrisation of
 a large class of hadronic correlators, including e.g.\ form factors
and the ordinary parton distribution functions. Thus GPDs provide a
solid formal basis
to connect information from various inclusive, semi-inclusive and
exclusive reactions in an efficient, unambiguous manner. Furthermore
they give access to physical quantities which cannot be directly
determined in experiments, like e.g.\ the orbital angular momentum
of quarks and gluons in a nucleon (for a chosen specific scheme) and
the spatial distribution of the energy or spin density of a fast
moving hadron in the transverse plane.
Since the structure of GPDs is rather complicated a
direct experimental
access is limited. Therefore, complementary
information channels have to be opened up.
One major source is lattice QCD~\cite{Hagler:2003jd,QCDSF,Gockeler:2004vx,Gockeler:2005aw,Gockeler:2005cj}.

Recently~\cite{Gockeler:2004xb}
we have calculated the
non-forward matrix elements 
in one-loop lattice perturbation theory for the
Wilson fermion action
needed for the renormalisation of the
second moments of GPDs.
From these results the renormalisation
factors for various representations of the hypercubic
group have been derived. 

Here we present some new results for operators with two covariant derivatives 
using the Sheikholeslami-Wohlert (clover)
action~\cite{Sheikholeslami:1985ij},
which leads to $O(a)$ improved quark-quark-gluon and quark-quark-gluon-gluon vertices in the Feynman rules. 
Since in current numerical simulations the operators for the second moment of GPDs are not improved, 
we ignore such a possible additional improvement.
Note that the $O(a)$ improvement for operators with one covariant derivative is known~\cite{Capitani:2000xi}.

We consider the Wilson gauge action and
clover fermions with
the fermionic action $S_{\rm SW,F}$~\cite{Sheikholeslami:1985ij} (for dimensionful massless
fermion fields $\psi(x)$)
\begin{eqnarray}
  S_{\rm SW,F}&=&
  4 r a^3 \sum_{x} \bar{\psi}(x)\psi(x)
  \\
  &-& \frac{a^3}{2}
  \sum_{x,\mu}\left[\bar{\psi}(x)(r-\gamma_\mu)U_{x,\mu} \psi(x+a \hat{\mu})
  \right.
  \nonumber
  \\
  &+& \left.\bar{\psi}(x+a \hat{\mu})(r+\gamma_\mu)U^\dagger_{x,\mu} \psi(x)\right]
  \nonumber
  \\
  &-& \frac{a^4\,g\,\csw}{4}
  \sum_{x,\mu,\nu}\,\bar{\psi}(x)\sigma_{\mu\nu}F^{\rm clover}_{\mu\nu} \psi(x)\,.
  \nonumber
\nonumber
\end{eqnarray}
Here $a$ denotes  the lattice spacing and the sums run over all lattice
sites $x$ and directions $\mu,\nu$
(all other indices are
suppressed).
$F^{\rm clover}_{\mu\nu}$ is the standard ``clover-leaf'' form of the
lattice field strength and $\sigma_{\mu\nu}=i/2[\gamma_\mu,\gamma_\nu]$.

In a perturbative calculation the operators to be investigated are
sandwiched between off-shell quark states with 4-momenta $p$ and $p'$.
Our calculations are performed in Feynman gauge, the final numbers
are presented for the Wilson parameter $r=1$, leaving the value
of $\csw$ free.

One-link quark operators with clover fermions have been discussed in~\cite{Capitani:2000xi}
for forward matrix elements.
The renormalisation constants found there for a
given representation of the
hypercubic group H(4) and charge conjugation parity $C$ can be used in the non-forward case as well. 
Additionally, in the case of GPDs ``transversity'' operators have to be taken into account.
We will collect here all results for completeness.

It is well known that operators with two or more covariant derivatives
may mix under H(4): the one- and higher-loop structures differ in general from that of the Born term and 
multiplicative renormalisation may get  lost.
In addition, for non-forward matrix elements
also operators with ordinary (external) derivatives can contribute 
making the mixing problem more complicated.
To find the possible candidates for mixing one has to define those operators which belong to the
same irreducible representation under H(4) and have the same charge conjugation parity.

We define renormalised operators $\mathcal{O}_i^{\cal S}$ by
\begin{equation}
{\cal O}_i ^{\cal S}(\mu) =  \sum_{k=1}^N Z_{ik}^S(a, \mu )\,{\cal O}_k (a)
\end{equation}
where $\mathcal{S}$ denotes the renormalisation scheme
and $N$ is the number of operators which mix in one-loop. 
$Z_{ik}^{\mathcal{S}}(a,\mu)$
are the renormalisation constants connecting the lattice operator $\mathcal{O}_k (a)$
with the renormalised operator $\mathcal{O}_i^{\mathcal{S}}(\mu)$ at scale $\mu$.
We present the renormalisation constants in the $\MS$ scheme following~\cite{Gockeler:2004xb}.

\section{OERATORS AND MIXING}
\label{OpMix}

We consider operators with up to two covariant symmetric lattice derivatives
$\Dlr=\Dr-\Dl$ and external ordinary derivatives $\partial$
needed for the chosen representations of interest for the first and second moment of GPDs.
The standard realisation of the covariant derivatives
acting to the right and to the left is used:
\begin{eqnarray}
 && \Dr_\mu \psi (x)  =\frac{1}{2a} \times
  \\
 &&  \Big[U_{x,\mu} \,
  \psi(x+a\hat{\mu}) 
  -U^\dagger_{x-a\hat{\mu},\mu} \, \psi(x-a\hat{\mu}) \Big]
  \,, \nonumber
  \\
&&  \bar{\psi} (x) \Dl_\mu 
=\frac{1}{2a} \times
 \\
&& \Big[
  \bar{\psi} (x+a\hat{\mu})\, U^\dagger_{x,\mu} 
  - \bar{\psi}(x-a\hat{\mu})
  \, U_{x-a\hat{\mu},\mu} \Big] \,.
  \nonumber
  \label{Dlr}
\end{eqnarray}
The external ordinary derivative is taken as
\begin{eqnarray}
 && \partial_\mu \left( \bar{\psi} \cdots \psi \right)\!(x)
  =\frac{1}{a} \times
  \\
  &&
  \Big[\left(\bar{\psi} \cdots \psi \right) (x+a\hat{\mu})
    -                \left(\bar{\psi} \cdots \psi \right)\!(x)
             \Big] \,.
  \nonumber
  \label{totderiv2}
\end{eqnarray}

The number of derivatives appearing in the operators is indicated by superscripts $D$ and $\partial$, respectively.
Quark operators with one derivative  are given by
\begin{eqnarray}
 \cO^D_{\mu\nu}&=& -\frac{i}{2}
  \bar\psi \gamma_\mu \Dlr_\nu \psi\,,
  \label{O11} \\
  \cO^{5,D}_{\mu\nu}&=& -\frac{i}{2}
  \bar\psi \gamma_\mu \gamma_5\Dlr_\nu \psi\,,
  \label{O12} \\
  \cO^{T,D}_{\mu\nu\omega}&=&-\frac{i}{2}
  \bar\psi [ \gamma_\mu , \gamma_\nu ] \Dlr_\omega \psi \,,
  \label{Oplow1} \\
  \cO^{T,\partial}_{\mu\nu\omega}&=&-\frac{i}{2}
  \partial_\omega
  \left( \bar\psi  [\gamma_\mu,\gamma_\nu ]  \psi \right)\,.
  \label{Oplow2}
\end{eqnarray}
The operator (\ref{Oplow1}) is a transversity operator antisymmetric in its first two indices
which is of interest for GPDs, operators (\ref{Oplow1}) and (\ref{Oplow2}) contribute as lower dimensional operators
to mixing in certain representations of the second moment of GPDs.

As operators with two derivatives we consider here
\begin{eqnarray}
  \cO_{\mu\nu\omega}^{DD}&=& -\frac{1}{4}
  \bar\psi \gamma_\mu \Dlr_\nu \Dlr_\omega\psi\,,
  \nonumber \\
  \cO_{\mu\nu\omega}^{\partial D}&=& -\frac{1}{4}
  \partial_\nu \left( \bar\psi \gamma_\mu \Dlr_\omega\psi \right) \,,
  \label{OpDD}
  \\
  \cO_{\mu\nu\omega}^{\partial \partial}&=& -\frac{1}{4}
  \partial_\nu \partial_\omega
  \left(\bar\psi \gamma_\mu \psi \right)
  \nonumber
  \,.
\end{eqnarray}
In addition, spin-dependent and ``transversity'' operators have to be considered
when discussing all possible representations.
They are roughly obtained by replacing $\gamma_\mu$ by $\gamma_\mu\gamma_5$ and
$\sigma_{\mu\tau}$, respectively. 

To define the various representations with given $C$
we use the following short-hand notations
\begin{eqnarray}
  \cO_{\cdots\{ \nu_1 \nu_2 \} }&=&\frac{1}{2} \left( \cO_{\cdots \nu_1\nu_2}+\cO_{\cdots \nu_2\nu_1} \right)
  \,,
  \nonumber\\
  \cO_{ \{ \nu_1\nu_2\nu_3 \} }&=& \frac{1}{6} \left(
  \cO_{\nu_1\nu_2\nu_3}+\cO_{\nu_1\nu_3\nu_2}+\cO_{\nu_2\nu_1\nu_3}\right.
  \nonumber
  \\
  &+& \left.\cO_{\nu_2\nu_3\nu_1}+\cO_{\nu_3\nu_1\nu_2}+\cO_{\nu_3\nu_2\nu_1}  \right)
  \,,
  \nonumber\\
  \cO_{\|\nu_1\nu_2\nu_3\| } &=& \cO_{\nu_1\nu_2\nu_3}-\cO_{\nu_1\nu_3\nu_2}
  \nonumber
  \\
  &+&\cO_{\nu_3\nu_1\nu_2}-\cO_{\nu_3\nu_2\nu_1}-2\,\cO_{\nu_2\nu_3\nu_1}
  \nonumber
  \\
  &+&2\,\cO_{\nu_2\nu_1\nu_3}
  \,,
  \nonumber\\
  \cO_{\langle\langle\nu_1\nu_2\nu_3\rangle\rangle } &=&
  \cO_{\nu_1\nu_2\nu_3}+\cO_{\nu_1\nu_3\nu_2}
  \nonumber
  \\
  &-&\cO_{\nu_3\nu_1\nu_2}-\cO_{\nu_3\nu_2\nu_1}
  \nonumber\,.
\end{eqnarray}
Let us denote an irreducible representation of the hypercubic group $H(4)$
by $\tau_k^{(l)}$ 
with dimension $l$ ($k$ labels  inequivalent representations of the same dimension)
and a given charge conjugation parity $C$ by $\pm1$.

For the first moments we choose the following representations
presented in Table~\ref{table1}
(for the notation and a detailed
discussion of the transformation under H(4) see~\cite{group}).
They are renormalised multiplicatively.
\begin{table}[!htb]
\vspace{-3mm}
\caption{Operators and their transformation under the hypercubic group.}
\label{table1}
\begin{tabular}{lcc}
  \hline
  Operator  & $\tau_k^{(l)}$ & $C$ \\
  \hline\\[-2ex]
  $\cO^D_{\{14\}}$                                                  & $\tau_3^{(6)}$ & $+1$\\ [0.7ex]
  $\cO^D_{44}-\frac{1}{3} \left(\cO^D_{11}+\cO^D_{22}+\cO^D_{33}\right) $    & $\tau_1^{(3)}$ & $+1$\\ [0.7ex]
  $\cO^{5,D}_{\{14\}}$& $\tau_4^{(6)}$ & $-1$\\ [0.7ex]
  $\cO^{5,D}_{44}-\frac{1}{3} \left(\cO^{5,D}_{11}+\cO^{5,D}_{22}+\cO^{5,D}_{33}\right) $ & $\tau_4^{(3)}$ & $-1$\\ [0.7ex]
  $\cO^{T,D}_{\langle\langle124\rangle\rangle}$ & $\tau_2^{(8)}$ & $+1$\\ [0.7ex]
  $\cO^{T,D}_{\langle\langle122\rangle\rangle}-\cO^{T,D}_{\langle\langle133\rangle\rangle}$ & $\tau_1^{(8)}$ & $+1$\\ [0.7ex]
  $\cO^{T,D}_{\|122\|}-\cO^{T,D}_{\|133\|}$ & $\tau_1^{(8)}$ & $-1$\\ [0.7ex]
  \hline
 \end{tabular}
\vspace{-0.5cm}
\end{table}

For the second moments we consider in this contribution the following mixing cases
( the details and additional operators will be presented elsewhere~\cite{future}):
\\
\underline{Representation $\tau_2^{(4)}$, $C=-1$} with operators
\begin{equation}
  \cO_{\{124\}}^{DD} \,, \, \cO_{\{124\}}^{\partial\partial} \,.
  \label{mixing1}
\end{equation}
\underline{Representation $\tau_1^{(8)}$, $C=-1$} with
\begin{eqnarray}
 \label{mixing2}
 &&\cO_1=\cO^{DD}_{\{114\}}-\frac{1}{2}
  \left(\cO^{DD}_{\{224\}}+\cO^{DD}_{\{334\}}\right)
  \,,
  \nonumber
  \\
  &&\cO_2=\cO^{\partial\partial}_{\{114\}}-\frac{1}{2} \left(
   \cO^{\partial\partial}_{\{224\}}
  +\cO^{\partial\partial}_{\{334\}}\right)
  \,,
  \nonumber
  \\
  &&\cO_3=\cO^{DD}_{\langle\langle 114\rangle\rangle}-\frac{1}{2}
  \left( \cO^{DD}_{\langle\langle224\rangle\rangle}+
         \cO^{DD}_{\langle\langle334\rangle\rangle}\right)
  \,,
  \nonumber
  \\
  &&\cO_4=\cO^{\partial\partial}_{\langle\langle 114 \rangle\rangle}-\frac{1}{2}
  \left(  \cO^{\partial\partial}_{\langle\langle 224 \rangle\rangle}+
          \cO^{\partial\partial}_{\langle\langle 334 \rangle\rangle}\right)
  \,,
  \nonumber
  \\
  &&\cO_5=\cO^{5,\partial D}_{||213||}
  \,, \quad
  \cO_6=\cO^{5,\partial D}_{\langle\langle213\rangle\rangle}
  \,,
  \label{O4}
  \\
  &&\cO_7=\cO^{5,DD}_{||213||}
  \,,
  \nonumber
  \\
  &&\cO_8=   \cO^{T,\partial}_{411}-
  \frac{1}{2}\left(\cO^{T,\partial}_{422}+\cO^{T,\partial}_{433} \right)
  \,.
  \nonumber
\end{eqnarray}

\section{ONE-LOOP CALCULATION}

We calculate the non-forward matrix elements of the operators in one-loop
lattice perturbation theory in the infinite volume limit
following Kawai et al.~\cite{Kawai:1980ja}.
Details of the computational procedure are given in~\cite{Gockeler:2004xb}.

In lattice momentum space operators with non-zero momentum transfer
$q$ are realised by applying the lattice momentum 
transfer $q$ at the lattice position $x$ or at the ``position 
centre'' $x+\frac{a}{2}\hat{\mu}$, e.g. for an operator with one covariant derivative we have 
the two possibilities
\begin{eqnarray}
  \label{DII}
& &\hspace{-5mm}\left(\bar{\psi}\Dlr_{\!\mu} \, \psi\right)\!(q) = \frac{1}{2a} \sum_x 
    \left\{
   {\rm e}^{i  q\cdot x}+{\rm e}^{i q \cdot(x+a\hat{\mu})}
    \atop
     2 {\rm e}^{i q \cdot(x+a\hat{\mu}/2)}
   \right\}
\times
\nonumber
 \\
 & &\hspace{-7mm}
 \left[\bar{\psi}(x)U_{x,\mu} \psi(x+a\hat{\mu})-
   \bar{\psi}(x+ a \hat{\mu})U^\dagger_{x,\mu} \psi(x)\right]
   \,.
\end{eqnarray}
Eq.~(\ref{DII}) basically defines the Feynman rules for the operators 
in lattice perturbation theory.
As an example we get for the operator $\cO^{DD}_{\mu\nu\omega}$ to order $O(g^0)$
($g$ is the bare gauge coupling):
\begin{eqnarray}
&&{\cal O}^{DD}
_{\mu\nu\omega}(p',p)  =  \bar\psi(p') \gamma_\mu \psi(p) \,
\times
\nonumber 
\\
&&
\frac{1}{a} \sin \frac{a (p+p')_\nu}{2}
 \, \frac{1}{a}   \sin \frac{a (p+p')_\omega}{2}
\times
\\
&&
\left\{
{
 \cos \frac{a (p-p')_\nu}{2} \, \cos \frac{a (p-p')_\omega}{2}
\atop
1}
\right\}
\,.
\nonumber
\end{eqnarray}
In the following sections we denote the upper/lower realisations by supercripts $I/II$.
 
\begin{figure}[!htb]
\vspace{2mm}
    \begin{center}
    \includegraphics[scale=0.5,clip=true]{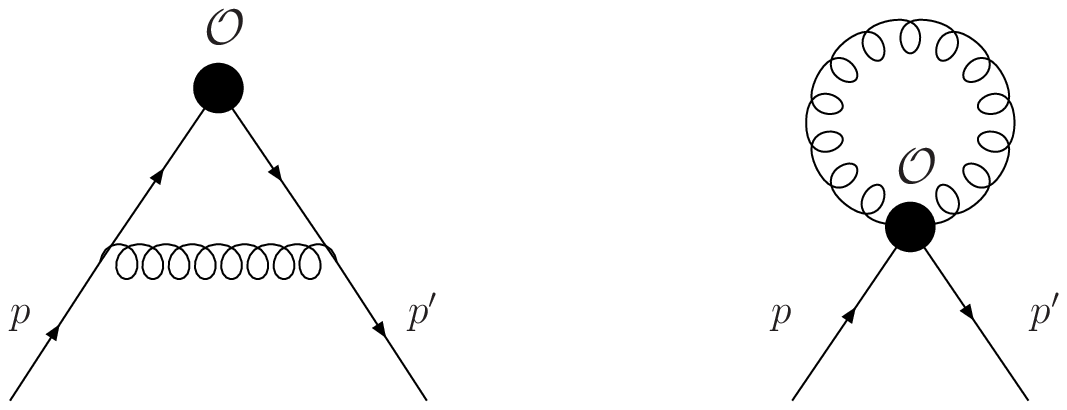} 
    \includegraphics[scale=0.5,clip=true]{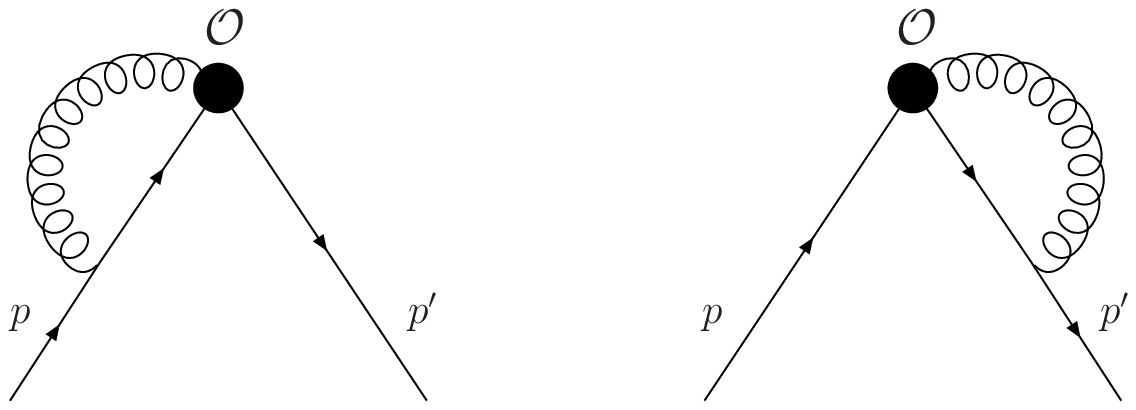} 
    \end{center}
\vspace{-8mm}
     \caption{Amputated Green function diagrams.}
     \label{fig:vertex}
\end{figure}

The contributing one-loop diagrams for the self energy and the amputated Green (vertex) functions are
shown in Figures~\ref{fig:selfenergy} and ~\ref{fig:vertex}
(filled black circles  indicate the place of the operator insertions):
\begin{figure}[!htb]
\vspace{-10mm}
    \begin{center}
    \includegraphics[scale=0.6,clip=true]{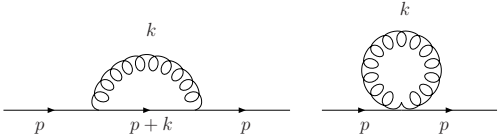} 
    \end{center}
 \vspace{-8mm}
     \caption{Quark self energy diagrams.}
     \label{fig:selfenergy}
\end{figure} 

\subsection{First Moment}

Since mixing is absent, we omit the matrix notation
for the renormalisation constants and use the general form
($C_F=4/3$, $g_R$ is the renormalised coupling)
\begin{equation}
  Z(a\mu) = 1 -
  \gRcf \left[\gamma\,\ln(a^2\mu^2) + B(\csw)\right]
  \label{Z}
\end{equation}
with the anomalous dimensions $\gamma=8/3$ for  the  first four operators  
and $\gamma=2$  for the  last  three  operators in Table~\ref{table1}.
For the operators of that Table
we get the finite contributions 
shown in Table~\ref{table2}.
\begin{table}[!htb]
\caption{Finite contributions $B(\csw)$ for the first moments.}
\label{table2}
\begin{tabular}{cc}
 \hline
  Representation & $B(\csw)$ \\
  \hline\\[-2ex]
  $\tau_3^{(6)}$ & $1.280 - 3.873\, \csw -0.678\, \csw^2 $\\ [0.7ex]
  $\tau_1^{(3)}$ & $2.562 - 3.970\, \csw -1.040\, \csw^2 $\\ [0.7ex]
  $\tau_4^{(6)}$ & $0.345 - 1.359\, \csw -1.893\, \csw^2 $\\ [0.7ex]
  $\tau_4^{(3)}$ & $0.167 - 1.249\, \csw -1.998\, \csw^2 $\\ [0.7ex]
  $\tau_2^{(8)}$ & $13.169 + 2.675\,\csw -1.494\, \csw^2 $\\ [0.7ex]
  $\tau_1^{(8)}$ ($C=\pm1$) & $12.804 + 2.624\,   \csw -1.430\,\csw^2 $\\ [0.7ex]
  \hline
  \end{tabular}
\vspace*{0.5cm}
\label{tabZ}
\end{table}

\subsection{Second Moment}

We present the matrix of renormalisation constants in the generic form
\begin{eqnarray}
  Z_{ij}^{(m)}(a\mu) &=&
  \nonumber
  \\
  & & \hspace{-2cm} \delta_{ij} -
  \gRcf \left[\gamma_{ij}\,\ln(a^2\mu^2)
  + B_{ij}^{(m)}(\csw) \right]
  \label{Zjk}
\end{eqnarray}
with
$$
B_{ij}^{(m)}(\csw)=B_{ij}^{(0,m)}+B_{ij}^{(1,m)}\,\csw+
		B_{ij}^{(2,m)}\,\csw^2\,.
$$
The superscript $(m)$ with $m=I,II$ distinguishes the
realisations I and II of the covariant derivatives~(\ref{DII}).

\vspace{0.5cm}
\underline{Representation  $\tau_2^{(4)}$, $C=-1$}
\vspace{0.2cm}

For this representation the operators (\ref{mixing1}) mix. The
corresponding $2 \times 2$-mixing matrices are
\begin{equation}
  \gamma_{jk}=\left(
  \begin{array}{rr}
     \frac{25}{6} & -\frac{5}{6}\\
       0          & 0
  \end{array}
  \right) \,,
  \label{andim1}
\end{equation}
\begin{eqnarray}
 && B_{jk}^{(I,II)}(\csw) =\left(
  \begin{array}{rr}
       -11.563    &  0.024\\
         0          & 20.618
  \end{array}
  \right)
  \nonumber \\
  && \hspace{19mm}+
\left(
  \begin{array}{rr}
       2.898     &  -0.255\\
         0          & 4.746
  \end{array}
  \right)\,\csw
  \nonumber
  \\
  && \hspace{19mm}-
\left(
  \begin{array}{rr}
       0.984     & \hspace{3mm} 0.016\\
         0          & 0.543
  \end{array}
  \right)\,\csw^2
   \,.
  \label{cjk}
\end{eqnarray}
In the matrix $B_{jk}^{(I,II)}$ the mixing between the operators
$\cO_{\{124\}}^{DD}$ and $\cO^{\partial\partial}_{\{124\}}$ is very small.
Thus it may be justified to neglect the mixing in practical applications.


\vspace{0.5cm}
\underline{Representation $\tau_1^{(8)}$, $C=-1$}
\vspace{0.2cm}

We consider the mixing (\ref{mixing2}) of the operators having the same dimension first.
These are the operators $\cO_1 \dots \cO_7$
in (\ref{mixing2}).
To one-loop accuracy the operator $\cO_7$ does not contribute and we have to consider the
following mixing set:
$$
  \{\cO_1,\cO_2, \cO_3,\cO_4,\cO_5,\cO_6\} \,.
$$
The
anomalous dimension matrix is
\begin{equation}
  \gamma_{jk} =\left( \begin{array}{rrrrrr}
 \frac{25}{6}&-\frac{5}{6}& 0          & 0          & 0          & 0\\
    0        & 0          & 0          & 0          & 0          & 0\\
    0        & 0          & \frac{7}{6}&-\frac{5}{6}&1           &-\frac{3}{2}\\
    0        & 0          & 0          & 0          & 0          & 0 \\
    0        & 0          & 0          & 0          & 2          & -2 \\
    0        & 0          & 0          & 0          & -\frac{2}{3}& \frac{2}{3}
  \end{array}
  \right)
  \label{andim2}
\end{equation}
and the finite  parts of the mixing matrix are given 
in Table~\ref{table3}.
(in cases of
doublets the upper number belongs to type I,
 the lower to type II of realisation of lattice
covariant derivative).
\begin{table*}[!htb]
\caption {Finite  parts of the mixing matrix for operators of representation $\tau_1^{(8)}$, $C=-1$.}
\label{table3}
\begin{eqnarray*}
  B_{jk}^{(0,I,II)} &=&\left( \begin{array}{rrrrrr}
 -12.127       & \left( \begin{array}{r}1.491\\-2.737 \end{array}\right)  & 0.368       &
 \left( \begin{array}{r}0.015\\0.994 \end{array}\right)       & 0.016    & 0.150\\[2ex]
    0          &  20.618      & 0           & 0           & 0        & 0\\[2ex]
  3.306        &  \left( \begin{array}{r}-8.015\\18.184 \end{array}\right)     &
  -14.852     & \left( \begin{array}{r}4.431\\-4.302 \end{array}\right)      & -0.928   & 0.738 \\[2ex]
    0          & 0            & 0           &  20.618     & 0        & 0 \\
 0             & 3.264        & 0           & 0           & 0.350    & 0.015 \\
 0             & 3.264        & 0           & 0           & 0.005    & 0.360
  \end{array}
  \right)
  \nonumber
  \\
\hspace{12mm}  B_{jk}^{(1,I,II)} &=&\left( \begin{array}{rrrrrr}
  2.922        & \left( \begin{array}{r}-0.213\\-0.686 \end{array}\right)       & -0.033
   & \left( \begin{array}{r}0.015\\0.173 \end{array}\right)       & -0.019   & 0.057\\[2ex]
    0          &  4.746       & 0           & 0           & 0        & 0\\[2ex]
  0.333        &  \left( \begin{array}{r}-0.766\\-0.055 \end{array}\right)      & 2.152
   & \left( \begin{array}{r}1.206\\0.970 \end{array}\right)       & -1.758   & 2.298 \\[2ex]
    0          & 0            & 0           & 4.746       & 0        & 0 \\
 0             & -1.441       & 0           & 0           & 1.648    & 0.866 \\
 0             & -1.441       & 0           & 0           & 0.289    & 2.225
  \end{array}
  \right)
\nonumber
\\
  B_{jk}^{(2,I,II)} &=&\left( \begin{array}{rrrrrr}
 -0.982        & \left( \begin{array}{r}-0.078\\-0.101 \end{array}\right)       & -0.029
 & \left( \begin{array}{r}0.035\\0.042 \end{array}\right)       & -0.001   & 0.007\\[2ex]
    0          & -0.543       & 0           & 0           & 0        & 0\\[2ex]
  0.371        & \left( \begin{array}{r}-0.551\\0.215 \end{array}\right)       & -1.707
  & \left( \begin{array}{r}0.371\\0.116 \end{array}\right)       & -0.443   & 0.103 \\[2ex]
    0          & 0            & 0           & -0.543      & 0        & 0 \\
 0             & 1.416        & 0           & 0           & -1.703   & 0.568 \\
 0             & 1.416        & 0           & 0           & 0.189    & -1.325
  \end{array}
  \right)
\nonumber
\end{eqnarray*}
\end{table*}

Using lattice perturbation theory to one-loop, $1/a$ terms may appear when calculating the matrix elements of the operators
with two covariant derivatives.
Such terms are potentially dangerous because of the power-law divergence in the 
continuum limit.
Considering the representation $\tau_2^{(4)}$, a potential  mixing is absent.
On the contrary, we get mixing for operator ${\cal O}_1$ of  $\tau_1^{(8)}$  with the
lower dimensional  operator $\cO_8$ given in (\ref{mixing2}).
The perturbative mixing result is   
\begin{eqnarray}
&&{\cal O}_1\big|_{1/a-{\rm part}} = \frac{g_R^2 C_F}{16 \pi^2} \,
\frac{1}{a} \, {\cal O}_8^{{\rm tree}} \times
\\
&&(-0.518+ 0.0832 \, \csw - 0.00983 \, \csw^2)
\,,
\nonumber
\end{eqnarray}
but a nonperturbative subtraction from the matrix element of 
${\cal O}_1$ is required to obtain reliable numbers.

\section{TADPOLE IMPROVEMENT AND SOME NUMERICAL EXAMPLES}

Since many results of (naive) lattice perturbation theory are in bad
agreement with their numerical counterparts, it has 
has been proposed~\cite{Lepage:1992xa} to rearrange the (naive) lattice perturbative series.
This rearrangement
is  performed using the variable $u_0$ (the mean field value of the link), e.g. 
defined from the measured value of the plaquette
at a given coupling
\begin{equation}
u_0 = \langle\frac{1}{3} {\rm Tr}U_\Box\rangle^{\frac{1}{4}} \,.
\end{equation}

In case of mixing the tadpole improvement procedure proceeds as follows.
By scaling the link variables $U_\mu$ with $u_0$
\begin{equation}
U_\mu(x) = u_0\,\left( \frac{U_\mu(x)}{u_0} \right) = u_0\,\overline{U}_\mu(x)
\nonumber
\end{equation}
the amputated Green function for operator $\cO$ with $n$
covariant derivatives $\Lambda^{(n)}_\mathcal{O}$ takes
the form
\begin{equation}
\Lambda^{(n)}_\mathcal{O} = u_0^{n} \Lambda^{(n)}_\mathcal{O}(\overline{U}_\mu(x))
\,.
\end{equation} 
$\Lambda^{(n)}_\mathcal{O}(\overline{U}_\mu(x))$ is expected to have a better
converging perturbative expansion. Up to order
$g^2$ we obtain for the Wilson gauge action, labelling the
operators by $i$ and the corresponding number of covariant derivatives
by $n_i$,
\begin{eqnarray}
\Lambda^{(n_i)}_i(\overline{U}_\mu(x))&=& \left(\frac{1}{u_0^{n_i}}\right)_{\rm pert}\,
\Lambda^{(n_i)}_{i,{\rm pert}}(U_\mu(x)) \nonumber \\
& =& \left(1+\frac{g^2 C_F}{16\,\pi^2} \, n_i \, \pi^2 +O(g^4) \right) \nonumber \\
& & \hspace{-2cm}\times \left(\Lambda^{(n_i,{\rm tree})}_{i} +
\frac{g^2 C_F}{16\,\pi^2} \sum_{k=1}^n w_{ik} \Lambda^{(n_k,{\rm tree})}_{k} \right.
\nonumber
\\ && \left. +O(g^4) \right)\,,
\label{TI1}
\end{eqnarray}
where the $w_{ik}$ denote the mixing weights. From (\ref{TI1}) it becomes
clear that in one-loop only the diagonal terms in the mixing matrix get a shift proportional
to $n_i\,\pi^2$.  An external ordinary derivative ($\partial$) does not provide a
factor of $u_0$. Taking into account the mean field value for the wave function
renormalisation constant for massless Wilson fermions
$
Z_{\psi,{\rm Wilson}}^{MF} = u_0
$
we get the tadpole improved matrix of renormalisation constants in the form
\begin{eqnarray}
Z_{ij}^{TI} &=& u_0^{1-n_i}\times
\nonumber
\\
&&\hspace{-1.0cm}\left(1-\frac{g^2 C_F}{16\,\pi^2}(n_i-1)\,\pi^2\, \delta_{ij} + O(g^4) \right)\,Z_{ij}\,.
\label{TI2}
\end{eqnarray}
Additionally, one has to replace the parameters
$g$ and $\csw$ by their {\it boosted} counterparts
\begin{eqnarray}
g_{TI}^2 \equiv g^2 \, u_0^{-4}\,, \quad
\csw^{TI} \equiv \csw \, u_0^{3}\,.
\label{TI3}
\end{eqnarray}
Putting (\ref{Zjk}), (\ref{TI2}) and (\ref{TI3}) together we obtain for the tadpole improved
renormalisation mixing matrix in one-loop order
\begin{eqnarray}
Z_{ij}^{TI,(m)} &=&  u_0^{1-n_i}\, \bigg(\delta_{ij} -
  \gTIcf  \times 
  \nonumber
  \\
 && \hspace{-5mm} \Big(\gamma_{ij}\,\ln(a^2\mu^2)
  + B_{ij}^{TI,(m)}(\csw^{TI})\Big)\bigg) 
\label{TI4}
\end{eqnarray}
with
\begin{equation}
B_{ij}^{TI,(m)}(\csw^{TI})= B_{ij}^{(m)}(\csw^{TI}) + (n_i-1)\pi^2\, \delta_{ij} \,.
\end{equation}

Let us demonstrate the effect of tadpole improvement by some numerical examples.
We choose  $a=1/\mu$,
$\beta=6$, $u_0=0.8778$ and $\csw=1+O(g^2)$~\cite{Capitani:2000xi}. 
For the first moments the only effect consists in replacing
$\csw$ by $\csw^{TI}$ and $g_R$ by $g_{TI}$ in (\ref{Z}) and in Table~\ref{tabZ}.
For the representation $\tau_3^{(6)}$  we get
\begin{equation}
Z = 1.028  \quad \rightarrow \quad Z^{TI} = 1.023\,.
\end{equation}

For the second moments 
we consider the simple mixing $\cO_{\{124\}}^{DD} \leftrightarrow \cO_{\{124\}}^{\partial\partial}$
(\ref{mixing1}) first. 
Without tadpole improvement we obtain the mixing matrix  
\begin{equation}
  Z_{ij} =\left( \begin{array}{rr}
  1.081        &   0.002 \\
    0          &   0.790
  \end{array}
  \right) \,.
\end{equation}
The tadpole improved result is
\begin{equation}
  Z_{ij}^{TI} =\left( \begin{array}{rr}
  1.142        &   0.002    \\
    0          &   0.707
  \end{array}
  \right) \,.
\end{equation}
It might be instructive to compare the one-loop corrections
for the renormalisation constants: $B^{(m)}_{ij}(\csw)$ for the
unimproved case (\ref{Zjk}) and $B_{ij}^{TI,(m)}(\csw^{TI})$
for the tadpole improved case (\ref{TI4}). 
We get 
\begin{equation}
  B_{ij} =\left( \begin{array}{rr}
  -9.649      &    -0.247    \\
    0          & 24.821
  \end{array}
  \right)
  \label{BUI1}
\end{equation}
and
\begin{equation}
  B_{ij}^{TI} =\left( \begin{array}{rr}
  -0.209       &   -0.177     \\
    0          &  12.035
  \end{array}
  \right) \,.
  \label{BTI1}
\end{equation}
We observe that in agreement with the improvement aims
the diagonal one-loop contributions are reduced.

For the representation  $\tau_1^{(8)}$, $C=-1$
with the mixing of operators $\mathcal{O}_1\dots\mathcal{O}_6$ 
we obtain for the unimproved/improved
mixing matrices (choosing $m=I$) 
the numbers given in Table~\ref{table4}.
\begin{table*}[!htb]
\caption{Unimproved and improved mixing matrices for the representation $\tau_1^{(8)}$, $C=-1$
at $a=1/\mu$, $\beta=6$, $u_0=0.8778$ and $\csw=1+O(g^2)$.}
\label{table4}
\begin{eqnarray*}
   Z_{ij}^{(I)}& =&\left( \begin{array}{rrrrrr}
  1.086        &   -0.010     & -0.003 &   -0.001    & 0.0   & -0.002\\
    0          &    0.790     & 0      & 0           & 0        & 0\\
  -0.034       &    0.079     & 1.122  & -0.051      & 0.026     & -0.027   \\
    0          & 0            & 0           & 0.790  & 0        & 0 \\
    0          & -0.027       & 0           & 0      & 0.998    & -0.012 \\
    0          & -0.027       & 0           & 0      & -0.004   & 0.989
  \end{array}
  \right) 
  \\
\hspace{20mm}   Z_{ij}^{TI,(I)} &=&\left( \begin{array}{rrrrrr}
  1.151        &   -0.021     & -0.005 &   -0.001    & 0.0   & -0.003\\
    0          &    0.707     & 0      & 0           & 0        & 0\\
  -0.060       &    0.142     & 1.209  & -0.088      & 0.037     & -0.038   \\
    0          & 0            & 0           & 0.707  & 0        & 0 \\
    0          & -0.042       & 0           & 0      & 0.990    & -0.012 \\
    0          & -0.042       & 0           & 0      & -0.004   & 0.982
  \end{array}
  \right) 
\end{eqnarray*}
\end{table*}

\section{SUMMARY}

Within the framework of lattice QCD with clover improved Wilson fermions and Wilson's
plaquette action for the gauge fields
we have calculated the one-loop quark matrix elements of operators needed for
the first two moments of GPDs and meson distribution amplitudes.
From these we have determined the matrices
of renormalisation and mixing coefficients in the $\MS$-scheme.

For the first moments of GPDs we can use the results from
the first moments of structure functions.
The results for  the second moments
extend the numbers obtained with  Wilson fermions~\cite{Gockeler:2004xb}.
The general conclusions concerning the mixing properties
remain unchanged. All  sets which consist of one operator with two
covariant derivatives $D$
and one operator with two external derivatives $\partial$ show very small mixing.
The set discussed here with seven potential candidates  (\ref{mixing2}) shows a 
more significant mixing.

Moreover, taking  $\cO_1$ from (\ref{mixing2})
as the operator to be
measured in a numerical simulation a mixing with a lower dimensional
operator $\cO_8$  appears.
This requires a nonperturbative subtraction for Wilson or clover fermions.
 
Using overlap fermions, such a mixing with a dangerous lower dimensional operator must be absent, since
the mixing operators are of different chirality.

\section*{Acknowledgements}

This work has been supported in part by
the EU Integrated Infrastructure Initiative Hadron Physics (I3HP) under
contract RII3-CT-2004-506078
and by the DFG under contract FOR 465 (Forschergruppe
Gitter-Hadronen-Ph\"anomenologie).

\end{document}